# Surface Wave-Based Radio Communication through Conductive Enclosures

Igor I. Smolyaninov [1)], Quirino Balzano [2)], and Dendy Young [3)]

*Abstract*—A surface wave antenna operating in the 2.4 GHz band and efficient for launching surface electromagnetic waves at metal/dielectric interfaces is presented. The antenna operation is based on the strong field enhancement at the antenna tip, which results in efficient excitation of surface waves propagating along nearby metal surfaces. Since surface electromagnetic waves may efficiently tunnel through deeply subwavelength channels from inner to outer metal/dielectric interface of a metal enclosure, this antenna is useful for broadband radio communication through various conductive enclosures, such as typical commercial Faraday cages.

## I. INTRODUCTION

It is very difficult to employ broadband radio signals for communication through conductive enclosures, such as underground tunnels, metal or partially metallic shipping containers, metallic test chambers, etc. Performance of conventional RF communication schemes in such situations is limited by the very small RF skin depth $\delta$, which may be calculated as:

$$\delta = \sqrt{\frac{1}{\pi \mu_0 \sigma \nu}} \quad , \tag{1}$$

where $\sigma$ is the medium conductivity and $\nu$ is the communication frequency [1], and by the Bethe's [2] expression for the transmission of a conventional TEM wave through a subwavelength aperture

$$T \propto \left(\frac{a}{\lambda}\right)^4 \quad , \tag{2}$$

where $a$ is the aperture size and $\lambda$ is the free space wavelength. As a result, conventional techniques of RF communication are impractical in situations where the walls of an enclosure are highly conductive, and the openings in the walls (if any) have deep subwavelength dimensions.

On the other hand, it is well established in the nanophotonics literature that efficient coupling to surface electromagnetic modes [3] (such as surface plasmon-polaritons at metal/dielectric interfaces) enables efficient light transmission through deeply subwavelength apertures [4-7]. Five to six orders of magnitude transmission enhancement has been observed in these experiments compared to the theoretical predictions based on Eq. (2). We have used a conceptually similar approach to demonstrate a battery-powered 2.45-GHz transmitting surface wave antenna, which is capable of sending video signals from inside a locked -90dB isolation commercial Faraday cage. This novel capability may be used for improving Wi-Fi connectivity in buildings and underground tunnels, as well as remote examination of metal and partially metal enclosures, such as shipping containers, metallic test chambers, etc. For example, very recently surface electromagnetic modes were used successfully to achieve broadband RF communication in seawater over distances, which considerably exceed the skin depth in seawater [8].

The principle of operation of the surface electromagnetic wave-based 2.4 GHz antenna is illustrated in Fig. 1. The electromagnetic field of the surface electromagnetic wave is partially longitudinal, which means that a good surface wave antenna needs to be placed in the vicinity of the metal surface, and it needs a strong field enhancement at its apex, which "pushes" charges along the metal surface. Below we will discuss the details of such surface wave antenna design and describe the operation of a battery-powered 2.4 GHz transmitter capable of sending high quality video signals through conductive enclosures.

---

[1)] I. I. Smolyaninov is with the Saltenna LLC, 1751 Pinnacle Drive, Suite 600 McLean VA 22102-4903 USA (e-mail: igor.smolyaninov@saltenna.com).

[2)] Q. Balzano is with the Electrical and Computer Engineering Department, University of Maryland, College Park, MD 20742 USA (e-mail: qbalzano@umd.edu).

[3)] D. Young is with the Saltenna LLC, 1751 Pinnacle Drive, Suite 600 McLean VA 22102-4903 USA (e-mail: dendy.young@saltenna.com).

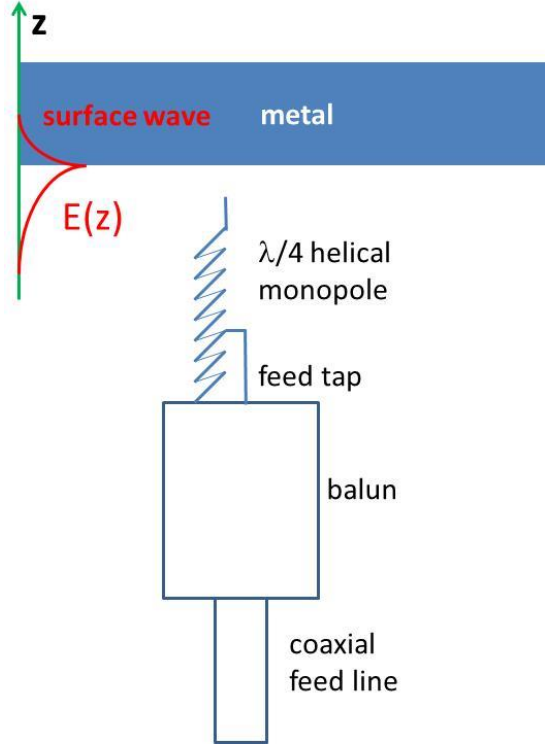

Fig. 1. Schematic geometry of a 2.4 GHz surface wave antenna design based on helical monopole shorted to its feed line outer conductor. The tip of the antenna is shown near a flat metal surface where it excites an omnidirectional surface electromagnetic wave . The electromagnetic field of the surface electromagnetic wave is partially longitudinal, which means that an efficient surface wave antenna needs a strong field enhancement at its apex, which "pushes" charges along the metal surface. A large charge accumulation at the tip is achieved using a low driving point impedance antenna. The antenna tip is also sharpened in order to take advantage of the lightning rod effect.

## II. Antenna Design and Discussion

An extensive review of the basic electromagnetic properties of surface electromagnetic waves (SEW) may be found in [3]. In particular, a SEW solution of Maxwell equations arises when the real part of the dielectric permittivity $\varepsilon$ changes sign across the interface. Unlike the conventional TEM waves in free space, this SEW propagating solution is partially longitudinal. It has a non-zero component of $E$ field along the propagation direction. A dielectric constant of a good metal is well described by the Drude model [9]:

$$\varepsilon_m(\omega) = 1 - \frac{\omega_p^2}{\omega^2 + i\omega\Gamma} \quad , \tag{3}$$

where $\omega_p$ is the plasma frequency and $\Gamma$ is the damping factor. The real part of $\varepsilon_m(\omega)$ is

$$\mathrm{Re}\,\varepsilon_m(\omega) = 1 - \frac{\omega_p^2}{\omega^2 + \Gamma^2} \tag{4}$$

At low frequencies

$$\mathrm{Re}\,\varepsilon_m(\omega) \approx 1 - \frac{\omega_p^2}{\Gamma^2}, \tag{5}$$

so the real part of $\varepsilon$ of typical metals is large and negative, while air has $\varepsilon=1$. Therefore, air-metal interfaces support the SEW modes.

As illustrated in Fig. 2, the SEW momentum is larger than the momentum of conventional TEM wave at the same frequency. Therefore, a plane TEM wave incident on a flat metal-dielectric interface cannot excite SEW due to momentum conservation. This is the main reason why SEWs are rarely seen in conventional RF experiments. However, in the configuration shown in Fig. 1, the metal surface which supports SEW propagation is placed within the near-field region of the antenna. In addition, placing the antenna near the plane metal wall breaks the conservation law for the momentum component parallel to the wall. Under such conditions various dipole sources are known to excite the SEW waves. However, compared to the conventional dipole antennas whose efficiency is compromised near metal surfaces, an additional advantage of our surface wave antenna design is that it is optimized for excitation of partially longitudinal waves. The strong field enhancement at the antenna apex "pushes" charges along the metal surface, leading to efficient excitation of SEWs.

It is also relatively straightforward to understand how SEWs facilitate transmission of electromagnetic energy through metal wall "imperfections", such as small fissures and gaps, which are always left in metal doors, even if the door locks look mechanically tight. Since both the internal and the external metal surfaces support SEW modes, these two surfaces act as weakly coupled SEW resonators, which may accumulate electromagnetic energy, and may also facilitate transmission of the electromagnetic energy across gaps between metal walls. While the small fissures and gaps cannot pass the conventional transverse electromagnetic (TEM) fields, SEW penetration is considerably higher in these situations. Indeed, for the conventional TEM wave the fraction of RF power transmitted through a small aperture a is typically defined by Eq.(2). which is negligible if $a<<\lambda$. On the other hand, for the SEW modes the same aperture in a metal wall is known to exhibit much higher transmission [4-7].

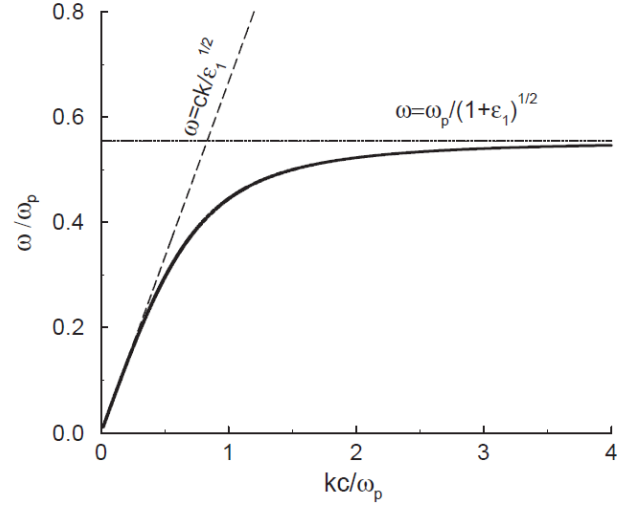

Fig. 2. Dispersion law of a surface electromagnetic wave (SEW) at a metal-dielectric interface [3]. $k=2\pi/\lambda_{SEW}$ is the SEW wave vector, $c$ is the speed of light, and $\varepsilon_1$ is the dielectric permittivity of the dielectric medium.

For example, Ebbesen et al. [4] have reported that "sharp peaks in transmission are observed at wavelengths as large as ten times the diameter of the holes. At these maxima, the transmission efficiency can exceed unity (when normalized to the area of the holes), which is orders of magnitude greater than predicted by standard aperture theory". The physical mechanism of this effect involves coupling of the SEW modes of the inner and outer metal interfaces, so that hybridized symmetric and anti-symmetric SEW modes are formed, which are strongly coupled to each other. In addition, SEW modes also exist in deeply subwavelength gaps, which also facilitate SEW transmission. As illustrated in Fig.2, SEW wavelength $\lambda_{SEW}$ may become much smaller than the wavelength of conventional TEM wave $\lambda$ in free space. As a result, application of the modified Bethe's equation to such a situation results in much higher transmission:

$$T \propto \left(\frac{a}{\lambda_{SEW}}\right)^4 \tag{6}$$

In this paper, following the analysis in [10], a $\lambda/4$ helical monopole (see Fig.1) resonant at 2.4 GHz, which is more efficient than a linear short monopole of the same length, has been used as a structure radiating SEWs, since the helical monopole design considerably enhances the electric field at its tip [11]. The charge at the tip is proportional to the divergence of the antenna current. A larger charge accumulation at the tip can be achieved using a low driving point impedance antenna, e.g. a helix. The antenna tip has also been sharpened in order to take advantage of the lightning rod effect, which increases the electric field near the tip, thus leading to enhanced SEW emission, as described in detail in [12]. This antenna requires careful tuning because the feed alters the electrical length of the antenna. A tuning procedure is illustrated in Figs. 3 and 4, which shows measurements of $S_{11}$ of the fabricated helical antennas resonant at 2.45 GHz, while they were nearly touching a large conductive copper plane. The length and the diameter of the helical antenna were selected for resonance at 2.45 GHz near the conductive plane. The finalized helical monopole was 3 cm long with 0.7 cm diameter. The tapping point for a 50 Ω match to a feeding coaxial line was determined using the setup of Fig. 3. Depending on the tapping point location, the radiative behavior of the helical antenna may be optimized for either 2D surface wave radiation, or 3D radiation into free space, as illustrated in Fig. 4.The final tuning of the antenna was performed by maximizing the received video signal outside a closed commercial Faraday chamber.

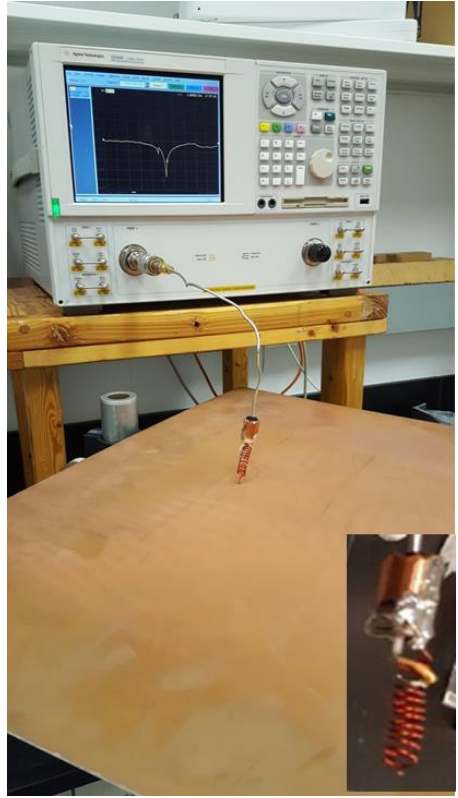

Fig. 3. Measurements of $S_{11}$ of the fabricated helical antenna resonant at 2.45GHz while nearly in contact with a large conductive copper plane. The inset shows a magnified photo of the constructed antenna.

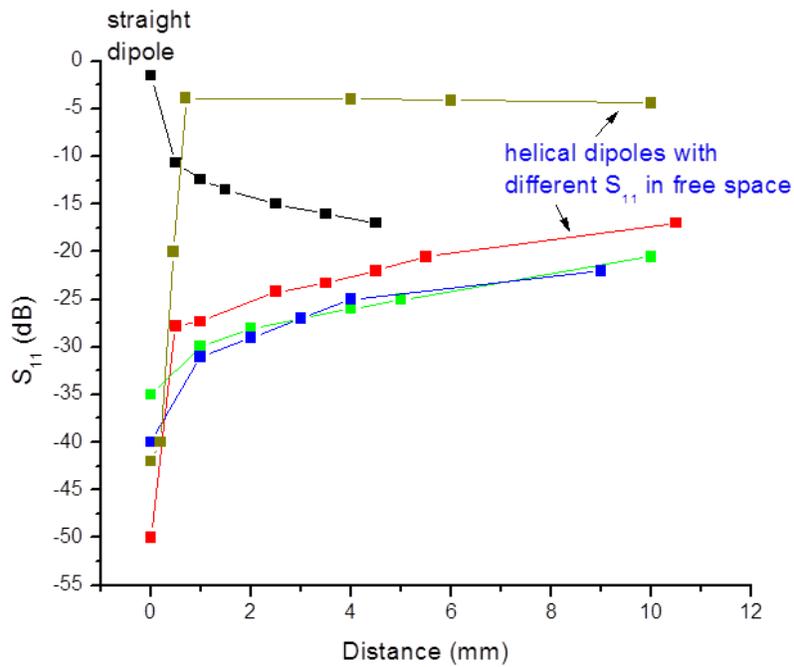

Fig. 4. Tuning of the fabricated helical antennas resonant at 2.45 GHz via measurements of $S_{11}$ as a function of distance from the large copper plane. The tuning parameter is the tapping point to a feeding coaxial line. Behavior of regular (straight) dipole antenna (see Fig. 6) optimized for radiation into free space is presented for a comparison.

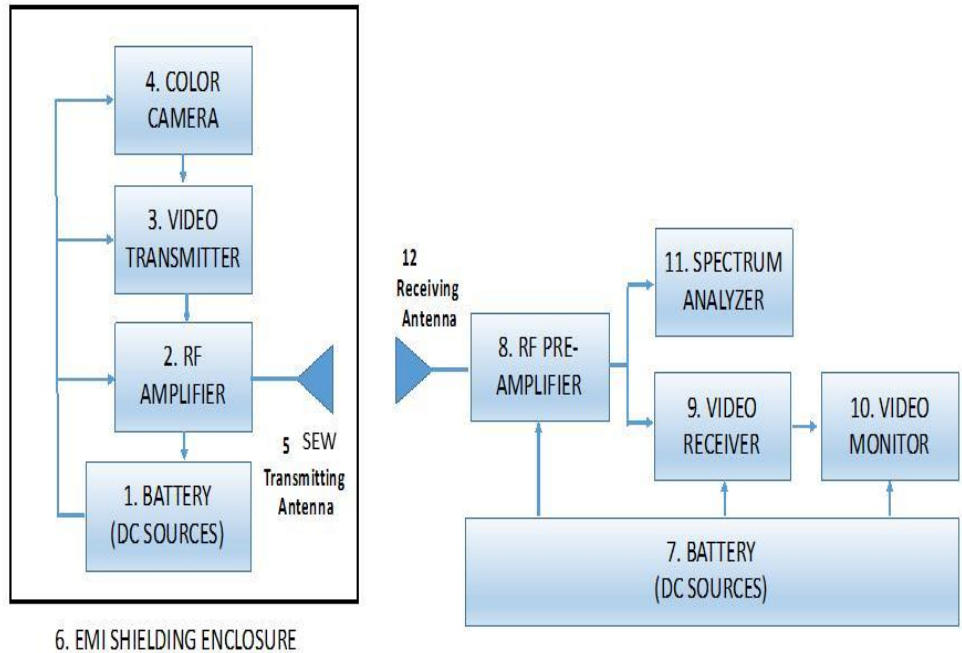

Fig. 5. Transmitting (left) and receiving (right) components of the SEW-based video transmission apparatus. The electromagnetic shielding enclosure (EMI) is rated at -90 dB isolation.

### III. MEASURED RESULTS

The operational performance of the surface wave antenna described above has been tested by transmitting Wi-Fi video signals through a commercial -90 dB isolation Faraday cage (JRE Test, model 0709), as illustrated schematically in Fig.5. Video signals generated from within the locked Faraday cage and transmitted live through free space without any cabling or connecting ground between the transmitter and receiver were received outside the enclosure at distances on the order of 10 to 100 cm and displayed on a live TV monitor. An amplified video signal generated inside the -90 dB isolation Faraday cage was fed via coaxial cable to the SEW transmitting antenna at 2.45 GHz. Outside the Faraday enclosure at a distance on the order of 10 to 100 cm, the video signal was received by either a conventional dipole antenna or a SEW antenna identical to the one of the transmitter. The received video signal was amplified, decoded, and displayed on a live video monitor. If necessary, during the measurements, the real-time visual output of the receive antenna was also decomposed into signal spectral amplitudes and displayed by a spectrum analyzer. The performance comparison of a conventional dipole antenna and the SEW antenna is illustrated in Figs. 6 and 7.

As expected, the conventional dipole antenna was not able to transmit video signal from inside the locked Faraday cage. The received signal fell below the -92 dBm sensitivity limit of the video receiver circuit, and was undetectable (video receiver RF Links, model VRX-24L, which uses NTSC/PAL encoding was used in this experiment). On the other hand, when the SEW transmit antenna was used inside the locked cage at the same transmit operating power, the live video connection was maintained. The typical received signals in this case (of the order of -55 dBm) are indicated in Fig. 8.

The SEW-mediated mechanism of 2.4 GHz video signal transmission through the locked Faraday cage has been verified by the transmitted signal measurements near the Faraday cage as a function of distance from the outside wall of the cage; results are shown in Fig.8. The exponential decay of the transmitted signal outside of the cage confirms its SEW character. We noted that the signal received farther away from the cage (in the far field zone) originates from the transmitted SEW field reaching the cage corners and scattering into the conventional propagating TEM fields. We have also verified that the transmitting SEW antenna efficiency strongly depends on the distance between the antenna tip and the inner wall of the Faraday cage, as illustrated by the red curve in Fig.8. In both cases, the antenna was tuned for maximum signal outside the Faraday cage.

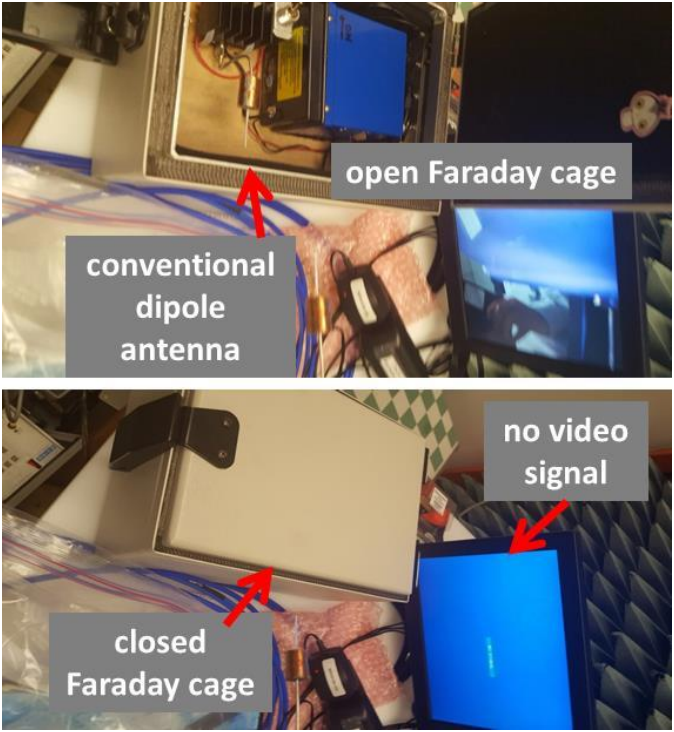

Fig. 6.  Conventional 2.4 GHz dipole antenna cannot transmit video signal from a locked -90dB Faraday cage.

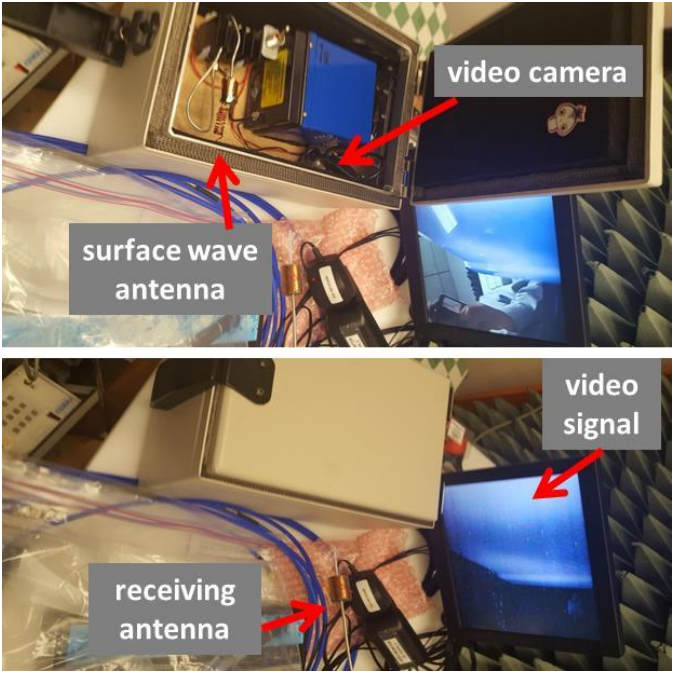

Fig. 7.  Surface wave antenna keeps transmitting video signal from a locked -90dB Faraday cage when operated at the same output power as the conventional dipole antenna shown in Fig.6.

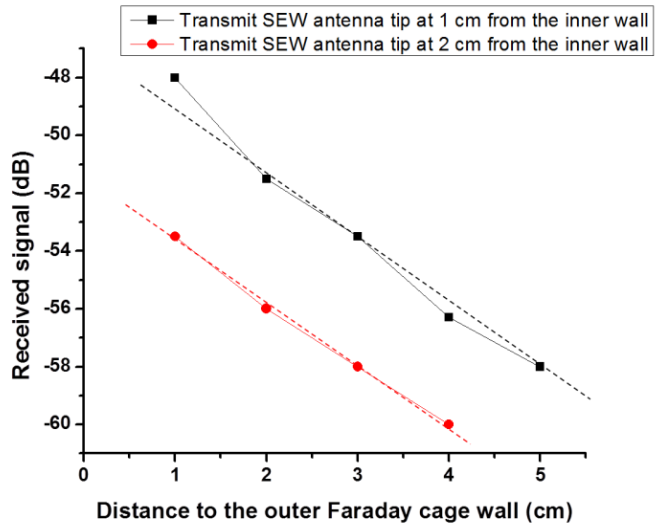

Figure 8. Transmitted signal measured near the Faraday cage as a function of distance from the outside wall of the cage in the configuration shown in Fig. 7.

We should note that the most obvious imperfection of the Faraday cage used in our experiments is its door. It is quite visible from the top photos in Figs. 6 and 7 that the door is protected from leakage by a metallic mesh, which is positioned in between the metallic double walls of the cage, and which has periodic millimeter-scale openings. SEWs are known to penetrate well through these kinds of periodic sub-wavelength defects [4-7].

We should also note that grounding the Faraday cage does not eliminate the video signal transmission from the inner SEW antenna. As was noted above, the SEW field is partially longitudinal. It has a nonzero electric field component parallel to the metal walls of the cage. Therefore, grounding the cage at some point(s) does not prevent SEW from propagating along the inner and outer walls of the cage, followed by scattering of the SEW propagating along the outer wall into the conventional TEM modes propagating in free space.

## IV. Conclusion

In this work we have demonstrated a successful realization of a 2.4 GHz band surface wave antenna that is efficient for launching surface electromagnetic waves at metal/dielectric interfaces. The antenna operation is based on the strong field enhancement at the antenna tip, which results in efficient excitation of surface waves propagating along nearby metal or highly conductive surfaces. Since surface electromagnetic waves may efficiently tunnel through deeply subwavelength channels from inner to outer metal/dielectric interface of a metal enclosure, this technique will be useful for broadband radio communication through various metal enclosures and Faraday cages. We anticipate that the demonstrated SEW antenna technique is scalable to other RF frequency ranges. We also believe that our work will result in considerable improvement of Wi-Fi connectivity in buildings, underground tunnels, as well as remote examination of metal and partially metal enclosures, such as shipping containers, metallic test chambers, etc. For example, in a very recent development [8], surface electromagnetic modes were used successfully to achieve broadband RF communication in seawater over distances, which considerably exceed the skin depth in seawater. Thus, the SEW-based antenna solutions together with new capabilities enabled by novel materials [13-20] will considerably enhance the range of THz and RF applications of surface electromagnetic waves.